\documentclass[aps, prd, showkeys, showpacs ]{revtex4}

\usepackage[small]{caption}
\usepackage{amsmath,amsfonts,amscd,amssymb,epsf, epsfig, mathrsfs,color}\usepackage{graphicx}

\usepackage{lscape,rotating}

\newcommand{\pa}{\partial}

\newcommand{\vep}{\varepsilon}

\begin{document}

 \title{Massive scalar Casimir interaction beyond proximity force approximation}

\author{L. P. Teo}
 \email{LeePeng.Teo@nottingham.edu.my}
\affiliation{Department of Applied Mathematics, Faculty of Engineering, University of Nottingham Malaysia Campus, Jalan Broga, 43500, Semenyih, Selangor Darul Ehsan, Malaysia.}
\begin{abstract}
Since massive scalar field plays an important role in theoretical physics, we consider the  interaction  between a sphere and a plate due to the vacuum fluctuation of a massive scalar field. We consider combinations of Dirichlet and Neumann boundary conditions.  There is a simple prescription to obtain the functional formulas for the Casimir interaction energies, known as TGTG formula, for the massive interactions from the massless interactions. From the TGTG formulas, we discuss how to compute the small separation asymptotic expansions of the Casimir interaction energies up to the next-to-leading order terms. Unlike the massless case, the results could not be expressed as simple algebraic expressions, but instead could only be expressed as infinite sums over some integrals. Nonetheless, it is easy to show that one can obtain the massless limits which agree with previously established results. We also show that the leading terms agree with that derive using proximity force approximation. The dependence of the leading order terms and the next-to-leading order terms on the mass of the scalar field is studied both numerically and analytically. In particular, we derive the small mass asymptotic expansions of these terms. Surprisingly, the small mass asymptotic expansions are quite complicated as they contain terms that are of odd powers in mass as well as logarithms of mass terms.
\end{abstract}
\pacs{03.70.+k, 12.20.Ds}
\keywords{  Casimir interaction, massive scalar field,   sphere-plate configuration, beyond proximity force approximation, Dirichlet boundary condition, Neumann boundary condition}

\maketitle
\section{Introduction}
The Casimir effect is an important quantum effect proposed about 70 years ago \cite{15}. The idea of the Casimir effect is very natural. The ground state energy of a quantum harmonic oscillator is not zero but is equal to $\hbar\omega/2$, where $\omega$ is the frequency of oscillator. A quantum field can be regarded as a superposition of infinitely many quantum harmonic oscillators. Hence it is natural to sum up the ground state energies of all the constituent oscillators, and the sum is known as the Casimir energy. In the presence of boundaries or external conditions, the Casimir energy would  create a nontrivial stress on the boundaries.

Although the idea is  simple, the actual computation of the Casimir energy is highly nontrivial because the sum is divergent. Various mathematical tools have been introduced to compute the Casimir energies of various systems, including zeta function method, exponential cut-off method, Green's function approach, etc. Various machineries have also been introduced to remove the divergencies. One setup that is known to produce renormalization independent result is the piston setup \cite{16}. In essence, the piston geometry gives rise to interaction between two parallel plates confined inside a long cylinder. In fact, after removing self-energies, the Casimir energy for the interaction between two compact objects is always finite.

Nevertheless, the computation of the Casimir interaction energy between two objects, where one or both of them is not planar, posted another major challenge to the field for a long time. Although Casimir experiments are performed for the sphere-plate interaction, it was much later when researchers successfully develop the necessary machineries to compute the Casimir interaction energy of the sphere-plate configuration. Before this, one relies on the proximity force approximation to obtain an approximation to the Casimir interaction force, under the assumption that when the separation between the objects is very small, a good approximation is obtained by dividing the pair of interacting surfaces into infinitely many pair of  parallel plates, and sum over the force of the latter. However, without the exact formula for the Casimir interaction energy, one cannot justify the validity range of the proximity force approximation.

A breakthrough to the problem appeared in 2006, where several groups of researchers employed different guises of multiple scattering method or the mode summation approach to obtain   explicit functional formulas for the Casimir interaction energies of the sphere-plate, sphere-sphere, cylinder-plate and cylinder-cylinder configurations \cite{17,18,19,20,21,22,23,24,25,26,27,28,29,30,31,32,33, 34, 35, 1}. These have made possible a more precise numerical or analytical study of the properties and strength of the Casimir interactions. One of the most interesting questions that is resolved is the validity of the proximity force approximation, and the correction to the proximity force approximation.

In the pioneering work \cite{24}, Bordag has developed a method to compute the small separation asymptotic expansion of the Casimir interaction energy between a cylinder and a plate up to the next-to-leading order term. This was later generalized to the sphere-plate, cylinder-cylinder and  sphere-sphere configurations \cite{37,38,39,40,41,42,43,44,45,46,47,48,49,50}.

Although the Casimir interaction due to massless scalar field is quite well-understood, so far no one has considered the Casimir interaction of massive scalar field between nontrivial objects. However, massive scalar field plays an important role in quantum field theory. The Casimir effect of massive scalar field has been studied in \cite{3,14,13,12,11,10,9,8,7,6,5,4,2} for the interaction between two plates or within a single spherical shell. The main objective of this work is to extend the studies of Casimir interaction between a sphere and a plate from massless scalar field to massive scalar field. We will concentrate on computing the correction to the proximity force approximation. For the sphere-plate interaction due to massless scalar field, it has been shown in \cite{37, 40} that, up to the next-to-leading order term,
\begin{align}\label{eq6_22_9}
E_{\text{Cas}}^{\text{XY}}\sim E_{\text{Cas}}^{\text{PFA}, \text{XY}}\left(1+\vartheta^{\text{XY}}\frac{d}{R} \right),
\end{align}
where X, Y $=$ D or N are the boundary conditions on the sphere and the plate,
$E_{\text{Cas}}^{\text{PFA}, \text{XY}}$ is the leading term that coincides with the proximity force approximation,
\begin{align*}
E_{\text{Cas}}^{\text{PFA}, \text{DD}}=E_{\text{Cas}}^{\text{PFA}, \text{NN}}=&-\frac{\hbar c \pi^3 R}{1440 d^2},\\
E_{\text{Cas}}^{\text{PFA}, \text{DN}}=E_{\text{Cas}}^{\text{PFA}, \text{ND}}=&\frac{7\hbar c \pi^3 R}{11520 d^2};
\end{align*}and $\vartheta^{\text{XY}}$ is a number that governs the correction to proximity force approximation, whose values are found to be
\begin{align*}
\vartheta^{\text{DD}}=&\vartheta^{\text{DN}}=\frac{1}{3},\\
\vartheta^{\text{ND}}=&\frac{1}{3}-\frac{160}{7\pi^2},\\
\vartheta^{\text{NN}}=&\frac{1}{3}-\frac{40}{\pi^2}.
\end{align*}
The main goal of this work is to find the corrections to these formulas when the mass of the scalar field is nonzero. Quite surprisingly, the formulas turn up to be quite complicated.

\section{Casimir interaction energy}\label{sec2}
The equation of motion of a massive scalar field is
\begin{align*}
\left(\frac{1}{c^2}\frac{\pa^2}{\pa t^2}-\frac{\pa^2}{\pa x^2}-\frac{\pa^2}{\pa y^2}-\frac{\pa^2}{\pa z^2}+\frac{m^2c^2}{\hbar^2}\right)\varphi=0.
\end{align*}
With
\begin{align*}
\varphi=\varphi(\mathbf{x})e^{-i\omega t},
\end{align*}we find that $\varphi(\mathbf{x})$ satisfies the equation
\begin{align}\label{eq6_19_1}
\left(\frac{\pa^2}{\pa x^2}+\frac{\pa^2}{\pa y^2}+\frac{\pa^2}{\pa z^2}\right)\varphi=-k^2\varphi,
\end{align}
where
\begin{align}\label{eq6_24_1}k=\sqrt{\frac{\omega^2}{c^2}-\frac{m^2c^2}{\hbar^2}}.\end{align}
Using the same method as in the massless case \cite{1}, we can derive the formula for the Casimir interaction energy. Since the details have been discussed in \cite{1}, we will not repeat them here. In fact, compare to the massless case, we find that $k$ is given by \eqref{eq6_24_1} instead of $k=\omega/c$ in the massless case.

 Assume that the sphere has radius $R$ and the distance between the sphere and the plate is $d$. Then the distance from the center of the sphere to the plate is $L=d+R$.
  The Casimir interaction energy is given by \begin{align}
E_{\text{Cas}}^{\text{XY}}=\frac{\hbar }{2\pi}\int_0^{\infty} d\xi \text{Tr}\ln\left(1-\mathbb{M}^{\text{XY}}\right),
\end{align}
where
\begin{equation}\label{eq6_19_2}\begin{split}
M_{lm,l'm'}^{\text{XY}}=   &(-1)^{\alpha_{\text{Y}}}\delta_{m,m'}\frac{\pi}{2}\sqrt{(2l+1)(2l'+1)\frac{(l-m)!(l'-m')!}{(l+m)!(l'+m')!}}
T_{l}^{\text{X}}\int_{0}^{\infty} d\theta\sinh\theta\\
&\times  P_l^m\left(\cosh\theta\right)
P_{l'}^m\left(\cosh\theta\right)e^{-2\kappa L\cosh\theta},\end{split}
\end{equation}X, Y $=$ D (Dirichlet) or N (Neumann) are the boundary conditions on the sphere and the plate, $\alpha_{\text{D}}=0$, $\alpha_{\text{N}}=1$,
\begin{align*}
T_{l}^{\text{D}}=&\frac{I_{l+\frac{1}{2}}(\kappa R)}{K_{l+\frac{1}{2}}(\kappa R)},\quad
T_{l}^{\text{N}}=\frac{-\frac{1}{2}I_{l+\frac{1}{2}}(\kappa R)+\kappa R I_{l+\frac{1}{2}}(\kappa R)}{-\frac{1}{2}K_{l+\frac{1}{2}}(\kappa R)+\kappa R K_{l+\frac{1}{2}}(\kappa R)},
\end{align*}and
\begin{align}\label{eq6_24_2}\kappa=\sqrt{\frac{\xi^2}{c^2}+\frac{m^2c^2}{\hbar^2}}.\end{align}
Here $I_{\nu}(z)$ and $K_{\nu}(z)$ are modified Bessel functions,  and $P_l^m (z)$ are associated Legendre functions \cite{51}.
This formula is almost the same as the formula in the massless case. We only have to replace
$\kappa=\xi/c$ in the formula for the massless case by the $\kappa$ given by \eqref{eq6_24_2}. One can use this same rule to obtain the formula of the Casimir interaction energy for massive scalar field from the corresponding formula for the massless scalar field.

Notice that as in the massless case, we can rewrite $M_{lm,l'm'}^{\text{XY}}$ as
\begin{align*}
M_{lm,l'm'}^{\text{XY}}=   &(-1)^{\alpha_{\text{Y}}}\delta_{m,m'}T_{l}^{\text{X}}\sqrt{\frac{\pi}{4\kappa L}}\sum_{l''=|l-l'|}^{l+l'}H_{ll';m}^{l''}K_{l''+\frac{1}{2}}(2\kappa L),
\end{align*}where $H_{ll';m}^{l''}$ can be expressed in terms of $3j$-symbols:
\begin{align*}
H_{ll';m}^{l''}=\sqrt{(2l+1)(2l'+1)}(2l''+1)\begin{pmatrix} l & l' & l''\\0&0&0\end{pmatrix}\begin{pmatrix} l & l' & l''\\m&-m&0\end{pmatrix}.
\end{align*}However, we will prefer the representation \eqref{eq6_19_2} since it is easier to compute the small separation asymptotic behavior using \eqref{eq6_19_2}.

\section{Small separation asymptotic expansion}
In this section, we derive the small separation asymptotic expansion of the Casimir interaction energy.

Introduce the dimensionless variables $$\vep=\frac{d}{R}, \quad \omega=\kappa R, \quad \mu=\frac{mc d}{\hbar},$$ we have
\begin{align}\label{eq6_19_3}
E_{\text{Cas}}^{\text{XY}}=&\frac{\hbar c }{2\pi R}\int_{\frac{\mu}{\vep}}^{\infty}d\omega \frac{\omega}{\sqrt{\omega^2-\left(\frac{\mu}{\vep}\right)^2}}\text{Tr}\ln\left(1-\mathbb{M}^{\text{XY}}\right).\end{align}
Now expanding the logartihm in \eqref{eq6_19_3} and replacing summations by integrations, we have
\begin{equation}\label{eq6_19_5}\begin{split}
E_{\text{Cas}}^{\text{XY}}\sim &-\frac{\hbar c }{2\pi R}\sum_{s=0}^{\infty}\frac{1}{s+1}\int_{0}^{\infty} dl\int_{\frac{\mu}{\vep}}^{\infty}d\omega \frac{\omega}{\sqrt{\omega^2-\left(\frac{\mu}{\vep}\right)^2}} \int_{-\infty}^{\infty} dm
\int_{-\infty}^{\infty} dn_1\ldots\int_{-\infty}^{\infty} dn_sM_{l_0m,l_1m}^{\text{XY}}\ldots M_{l_sm,l_0m}^{\text{XY}}.
\end{split}\end{equation}Here we have reparametrized the variables:
$$l_0=l,\hspace{1cm}l_i=l+n_i,\quad 1\leq i\leq s.$$
Now using the integral representation (see \cite{51, 44})
\begin{align*}
P_l^m(\cosh\theta) =&(-1)^m \frac{(l+m)!}{\pi  }\sum_{k=0}^l\frac{1}{k!(l-k)!}e^{(l-2k)\theta}\int_{-\frac{\pi}{2}}^{\frac{\pi}{2}}d\varphi  \cos^{2l-2k}\varphi \sin^{2k}\varphi e^{2im\varphi},\end{align*}
we find that
\begin{align*}
M_{l_im,l_{i+1}m}^{\text{XY}}=   &(-1)^{\alpha_{\text{Y}}} \frac{1}{2\pi}\sqrt{(2l_i+1)(2l_{i+1}+1) (l_i-m)!(l_{i+1}-m)! (l_i+m)!(l_{i+1}+m)! }\;
T_{l_i}^{\text{X}}
\\
&\times  \sum_{k=0}^{l_i}\frac{1}{k!(l_i-k)!} \sum_{k'=0}^{l_{i+1}}\frac{1}{k'!(l_{i+1}-k')!}e^{(l_i+l_{i+1}-2k-2k')\theta}\int_{0}^{\infty} d\theta\sinh\theta e^{-2\omega(1+\vep)\cosh\theta}\\&\times \int_{-\frac{\pi}{2}}^{\frac{\pi}{2}}d\varphi  \cos^{2l_i-2k}\varphi \sin^{2k}\varphi e^{2im\varphi}\int_{-\frac{\pi}{2}}^{\frac{\pi}{2}}d\varphi'  \cos^{2l_{i+1}-2k'}\varphi' \sin^{2k'}\varphi' e^{2im\varphi'}.
\end{align*}
Let
\begin{align*}
\tau=\frac{l}{\sqrt{\omega^2+l^2}},
\end{align*}
so that
\begin{align*}
\omega=\frac{l\sqrt{1-\tau^2}}{\tau}.
\end{align*}
Also, we need to make a change of variables
$$\theta=\theta_0+\tilde{\theta},$$ where
$$\sin\theta_0=\frac{\tau}{\sqrt{1-\tau^2}}.$$
Now we can expand  every term in small $\vep$, keeping in mind that $l$ has order $\vep^{-1}$, $n_i$ and $m$  have order $\vep^{-\frac{1}{2}}$. In the following, we denote by $\mathcal{X}_{i,1}$ and $\mathcal{X}_{i,2}$ terms of order $\sqrt{\vep}$ and $\vep$. With the help of symbolic mathematics softwares, we find that
\begin{align*}
\int_{-\frac{\pi}{2}}^{\frac{\pi}{2}}d\varphi  \cos^{2l_i-2k}\varphi \sin^{2k}\varphi e^{2im\varphi}\sim &\int_{-\frac{\pi}{2}}^{\frac{\pi}{2}}d\varphi
\varphi^{2k}\left(1-\frac{\varphi^2}{6}\right)^{2k}\exp\left(-(2l+2n_i-2k)\left(\frac{\varphi^2}{2}+\frac{\varphi^4}{12}\right)\right)e^{2im\varphi}\\
\sim &\frac{1}{l^{k+\frac{1}{2}}}\int_{-\infty}^{\infty}d\tilde{\varphi}
\tilde{\varphi}^{2k}\left(1+\mathcal{B}_{2,i}\right) \exp\left(-\tilde{\varphi}^2+2im\frac{\tilde{\varphi}}{\sqrt{l}}+\mathcal{A}_{i,1}+\mathcal{A}_{i,2}\right).
\end{align*} Similarly, we have
\begin{align*}
\int_{-\frac{\pi}{2}}^{\frac{\pi}{2}}d\varphi'  \cos^{2l_{i+1}-2k'}\varphi' \sin^{2k'}\varphi' e^{2im\varphi'} \sim &\frac{1}{l^{k'+\frac{1}{2}}}\int_{-\infty}^{\infty}d\tilde{\varphi}'
\tilde{\varphi}^{\prime 2k'}\left(1+\mathcal{D}_{2,i}\right) \exp\left(-\tilde{\varphi}^{\prime 2}+2im\frac{\tilde{\varphi}'}{\sqrt{l}}+\mathcal{C}_{i,1}+\mathcal{C}_{i,2}\right).
\end{align*}
On the other hand,
\begin{align*}
\sinh\theta=&\frac{\tau}{\sqrt{1-\tau^2}}\left(1+\frac{\tilde{\theta}}{\tau}+\frac{\tilde{\theta}^2}{2}\right)=\frac{\tau}{\sqrt{1-\tau^2}}\left(1+\mathcal{E}_{i,1}
+\mathcal{E}_{i,2}\right),
\end{align*}
\begin{align*}
&e^{-2\omega(1+\vep)\cosh\theta}e^{(l_i+l_{i+1}-2k-2k')\theta}\\\sim & \left(\frac{1+\tau}{1-\tau}\right)^{l+\frac{n_i+n_{i+1}}{2}-k-k'}\exp\left(
(2l+n_i+n_{i+1}-2k-2k')\tilde{\theta} -\frac{2l}{\tau}(1+\vep)\left(1+\tau\tilde{\theta}+\frac{\tilde{\theta}^2}{2}+\tau\frac{\tilde{\theta}^3}{6}+\frac{\tilde{\theta}^4}{24}\right)\right)\\
\sim & \left(\frac{1+\tau}{1-\tau}\right)^{l+\frac{n_i+n_{i+1}}{2}-k-k'}\exp\left(
 -\frac{2l}{\tau}-\frac{2l\vep}{\tau}-\frac{l\tilde{\theta}^2}{\tau} +(n_i+n_{i+1})\tilde{\theta}+\mathcal{F}_{i,1}+\mathcal{F}_{i,2}\right),
\end{align*}
\begin{align*}
\sqrt{(2l_i+1)(2l_{i+1}+1) }=&2l\left(1+\mathcal{G}_{i,1}+\mathcal{G}_{i,2}\right),
\end{align*}
\begin{align*}
\frac{\sqrt{(l_i-m)!(l_{i+1}-m)! (l_i+m)!(l_{i+1}+m)! }}{(l_i-k)!(l_{i+1}-k')!}
\sim l^{k+k'}\exp\left(\frac{m^2}{l}+\mathcal{H}_{i,1}+\mathcal{H}_{i,2}\right).
\end{align*}
Using Debye asymptotic expansions for modified Bessel functions \cite{52}:
 \begin{equation}\label{eq3_26_6}\begin{split}
I_{\nu}(\nu z)\sim & \frac{1}{\sqrt{2\pi \nu}}\frac{e^{\nu\eta(z)}}{(1+z^2)^{\frac{1}{4}}}\left(1+\frac{u_1(t(z))}{\nu}+\ldots\right),\\
K_{\nu}(\nu z)\sim &\sqrt{\frac{\pi}{ 2 \nu}}\frac{e^{-\nu\eta(z)}}{(1+z^2)^{\frac{1}{4}}}\left(1-\frac{u_1(t(z))}{\nu}+\ldots\right),\\
I_{\nu}'(\nu z)\sim & \frac{1}{\sqrt{2\pi \nu}}\frac{e^{\nu\eta(z)}(1+z^2)^{\frac{1}{4}}}{z}\left(1+\frac{v_1(t(z))}{\nu}+\ldots\right),\\
K_{\nu}'(\nu z)\sim &-\sqrt{\frac{\pi}{ 2 \nu}}\frac{e^{-\nu\eta(z)}(1+z^2)^{\frac{1}{4}}}{z}\left(1-\frac{v_1(t(z))}{\nu}+\ldots\right),
\end{split}\end{equation}where
\begin{equation}\begin{split}
\eta(z)=&\sqrt{1+z^2}+\log\frac{z}{1+\sqrt{1+z^2}},\hspace{1cm}
t(z)=\frac{1}{\sqrt{1+z^2}},\\
u_1(t)=&\frac{t}{8}-\frac{5t^3}{24},\hspace{1cm}
v_1(t)=-\frac{3t}{8}+\frac{7t^3}{24},
\end{split}\end{equation}we find that
\begin{align*}
T_{l_i}^{\text{X}}\sim & (-1)^{\alpha_{\text{X}}}C^{n_i-n_{i+1}}\frac{1}{\pi}\left(\frac{1+\tau}{1-\tau}\right)^{-l-\frac{n_i+n_{i+1}}{2}-\frac{1}{2}}\exp\left(\frac{2l}{\tau}
-\frac{\tau}{2l}\left(n_i^2+n_{i+1}^2\right)+\mathcal{N}_{i,1}+\mathcal{N}_{i,2}\right)\left(1+\mathcal{M}^{\text{X}}\right).
\end{align*}
Here
\begin{align*}
\mathcal{M}^{\text{D}}=\frac{2}{l}\left(\frac{\tau}{8}-\frac{5\tau^3}{24}\right),\quad \mathcal{M}^{\text{N}}=\frac{2}{l}\left(-\frac{7\tau}{8}+\frac{7\tau^3}{24}\right)
\end{align*} are terms of order $\vep$.

Collate everything, sum over $k$ and $k'$ from 0 to infinity using
\begin{align*}
\sum_{k=0}^{\infty} \frac{v^k}{k!}=e^v,\quad \sum_{k=0}^{\infty} k\frac{v^k}{k!}=ve^v,\quad \sum_{k=0}^{\infty}k^2 \frac{v^k}{k!}=(v^2+v)e^v,
\end{align*}
 and then integrate over $\tilde{\varphi}$, $\tilde{\varphi}'$ and $\tilde{\theta}$, we find that
\begin{align*}
M_{l_im,l_{i+1}m}^{\text{XY}}=&C^{n_i-n_{i+1}}(-1)^{\alpha_{\text{X}}+\alpha_{\text{Y}}}\frac{1}{2}\sqrt{\frac{\tau}{\pi l}}   \left(1+\mathcal{O}_{i,1}
+\mathcal{O}_{i,2}\right) \exp\left(
 -\frac{2l\vep}{\tau} -\frac{m^2}{l\tau}-\frac{\tau}{4l}\left(n_i-n_{i+1}\right)^2 \right)\left(1+\mathcal{M}^{\text{X}}\right).
\end{align*} Substitute into \eqref{eq6_19_5}, we find that
\begin{equation}\label{eq6_19_6}\begin{split}
E_{\text{Cas}}^{\text{XY}}
\sim &-\frac{\hbar c }{2\pi R}\sum_{s=0}^{\infty}\frac{(-1)^{(s+1)(\alpha_{\text{X}}+\alpha_{\text{Y}})}}{s+1}\frac{1}{2^{s+1}\pi^{\frac{s+1}{2}}}\int_{\frac{\mu}{\vep}}^{\infty}d\omega \frac{\omega}{\sqrt{\omega^2-\left(\frac{\mu}{\vep}\right)^2}} \int_{0}^{\infty} dl l^{-\frac{s+1}{2}}\tau^{\frac{s+1}{2}}\int_{-\infty}^{\infty} dm
\int_{-\infty}^{\infty} dn_1\ldots\int_{-\infty}^{\infty} dn_s \\
&\times \left(1+\sum_{i=0}^{s} \mathcal{O}_{i,1}+\sum_{i=0}^{s-1}\sum_{j=i+1}^s \mathcal{O}_{i,1}\mathcal{O}_{j,1}+\sum_{i=0}^s \mathcal{O}_{i,2}+(s+1)\mathcal{M}^{\text{X}}\right) \\&\times \exp\left(
 -\frac{2l(s+1)\vep}{\tau} -\frac{m^2(s+1)}{l\tau}-\frac{\tau}{4l}\sum_{i=0}^s\left(n_i-n_{i+1}\right)^2 \right).\end{split}\end{equation}
 After integrating over $n_i$ and $m$, this gives
 \begin{align*}
 E_{\text{Cas}}^{\text{XY}}
\sim & -\frac{\hbar c }{4\pi R}\sum_{s=0}^{\infty}\frac{(-1)^{(s+1)(\alpha_{\text{X}}+\alpha_{\text{Y}})}}{(s+1)^{2}}  \int_{0}^{\infty} dl\; \int_{\frac{\mu}{\vep}}^{\infty}d\omega \frac{\omega}{\sqrt{\omega^2-\left(\frac{\mu}{\vep}\right)^2}} \frac{l}{\sqrt{\omega^2+l^2}}
   \left(1+\mathcal{Q}^{\text{X}}\right) \exp\left(
 - 2 (s+1)\vep\sqrt{\omega^2+l^2}   \right).
\end{align*}$\mathcal{Q}^{\text{X}}$ is a term of order $\vep$.  The term
$$\sum_{i=0}^{s} \mathcal{O}_{i,1}$$in \eqref{eq6_19_6} does not give any contribution since it is odd in one of the $n_i$'s.
Now, making the change of variables
$$v=\vep l,\quad t=\vep\sqrt{\omega^2+l^2},$$we find that
\begin{align*}
 E_{\text{Cas}}^{\text{XY}}
\sim & -\frac{\hbar c R}{4\pi d^2}\sum_{s=0}^{\infty}\frac{(-1)^{(s+1)(\alpha_{\text{X}}+\alpha_{\text{Y}})}}{(s+1)^{2}}  \int_{0}^{\infty} dv\; v\int_{\sqrt{v^2+\mu^2}}^{\infty}  \frac{dt}{\sqrt{t^2-v^2-\mu^2}}
   \left(1+\mathcal{Q}^{\text{X}}\right) \exp\left(
 - 2 (s+1)t  \right)\\
 =&-\frac{\hbar c R}{4\pi d^2}\sum_{s=0}^{\infty}\frac{(-1)^{(s+1)(\alpha_{\text{X}}+\alpha_{\text{Y}})}}{(s+1)^{2}}  \int_{\mu}^{\infty} dt\int_{0}^{\sqrt{t^2-\mu^2}} dv \frac{v}{\sqrt{t^2- \mu^2-v^2}}
   \left(1+\mathcal{Q}^{\text{X}}\right) \exp\left(
 - 2 (s+1)t  \right).
\end{align*}
Using
\begin{align*}
\int_0^adx\frac{x^{b}}{\sqrt{a^2-x^2}} =& \frac{\sqrt{\pi}a^b}{2}\frac{\Gamma\left(\frac{b+1}{2}\right)}{\Gamma\left(\frac{b+2}{2}\right)},
\end{align*}
we find that
\begin{align*}
E_{\text{Cas}}^{\text{XY}}
\sim &  -\frac{\hbar c R}{4\pi d^2}\sum_{s=0}^{\infty}\frac{(-1)^{(s+1)(\alpha_{\text{X}}+\alpha_{\text{Y}})}}{(s+1)^{2}}   \int_{\mu}^{\infty} dt \sqrt{t^2 -\mu^2}
   \left(1+\mathcal{S}^{\text{X}} \right) \exp\left(
 - 2 (s+1) t  \right),
\end{align*}where
\begin{align*}
\mathcal{S}^{\text{D}}=&\vep\left\{\frac{1}{ \sqrt{t^2-\mu^2}}\frac{\pi}{8t^2}\left(-2(s+1)t^3+t^2+2\mu^2(s+1)t+\mu^2\right)\right.\\
&+\frac{2 \left((s+1)^3+2(s+1)\right)}{9}\left(t-\frac{\mu^2}{t}\right)-\frac{1}{9}\left(4(s+1)^2-1\right)
+\frac{1}{9  t}\left((s+1)-\frac{1}{s+1}\right)\\&\left.-\frac{2 \mu^2}{9t^2}\left((s+1)^2+2\right)+\frac{2\mu^2}{9 t^3}\left((s+1)-\frac{1}{s+1}\right)\right\},\\
\mathcal{S}^{\text{N}}=&\mathcal{S}^{\text{D}}+\vep\left\{-\frac{4}{3t}(s+1)-\frac{2\mu^2}{3t^3}(s+1)\right\}.
\end{align*}
Notice that
\begin{align*}
&\int_{\mu}^{\infty} dt  \frac{\pi}{8t^2}\left(-2(s+1)t^3+t^2+2\mu^2(s+1)t+\mu^2\right)
    \exp\left(
 - 2 (s+1) t  \right)\\=&\int_{\mu}^{\infty} dt  \frac{\pi}{8 }\frac{d}{dt}\left\{\left(t-\frac{\mu^2}{t}\right)
    \exp\left(
 - 2 (s+1) t  \right)\right\}\\
 =&\left[\left(t-\frac{\mu^2}{t}\right)
    \exp\left(
 - 2 (s+1) t  \right)\right]_{\mu}^{\infty}=0.
\end{align*}
On the other hand,
\begin{align*}
&\left\{\frac{2 \left((s+1)^3+2(s+1)\right)}{9}\left(t-\frac{\mu^2}{t}\right)-\frac{1}{9}\left(4(s+1)^2-1\right)
+\frac{1}{9  t}\left((s+1)-\frac{1}{s+1}\right)\right.\\&\left.-\frac{2 \mu^2}{9t^2}\left((s+1)^2+2\right)+\frac{2\mu^2}{9 t^3}\left((s+1)-\frac{1}{s+1}\right)\right\}\sqrt{t^2-\mu^2}e^{-2(s+1)t}\\
=&\frac{d}{dt}\left\{ -\frac{(s+1)^2}{9} \left(t-\frac{\mu^2}{t}\right)\sqrt{t^2-\mu^2}e^{-2(s+1)t}+\frac{(s+1)}{9}\left(1-\frac{\mu^2}{t^2}\right)\sqrt{t^2-\mu^2}e^{-2(s+1)t}\right.
\\&\left. -\frac{2}{9} \left(t-\frac{\mu^2}{t}\right)\sqrt{t^2-\mu^2}e^{-2(s+1)t}-\frac{1}{9(s+1)}\left(1-\frac{\mu^2}{t^2}\right)\sqrt{t^2-\mu^2}e^{-2(s+1)t}\right\}\\
&+\left\{\frac{1}{3}-\frac{1}{3}\frac{\mu^2}{t^2}(s+1)^2 \right\}\sqrt{t^2-\mu^2}e^{-2(s+1)t}.
\end{align*}
The total derivative part vanishes when $t=\mu$ and $t\rightarrow\infty$. Hence, we find that
\begin{align*}
E_{\text{Cas}}^{\text{XY}}
\sim &  -\frac{\hbar c R}{4\pi d^2}\sum_{s=0}^{\infty}\frac{(-1)^{(s+1)(\alpha_{\text{X}}+\alpha_{\text{Y}})}}{(s+1)^{2}}   \int_{\mu}^{\infty} dt \sqrt{t^2 -\mu^2}
   \left(1+\mathcal{T}^{\text{X}} \right) \exp\left(
 - 2 (s+1) t  \right),
\end{align*}
where
\begin{align*}
\mathcal{T}^{\text{D}}=&\vep\left\{\frac{1}{3}-\frac{1}{3}\frac{\mu^2}{t^2}(s+1)^2\right\},\\
\mathcal{T}^{\text{N}}=&\vep\left\{\frac{1}{3}-\frac{1}{3}\frac{\mu^2}{t^2}(s+1)^2-\frac{4}{3t}(s+1)-\frac{2\mu^2}{3t^3}(s+1)\right\}.
\end{align*}
The leading term of the Casimir interaction energy is
\begin{equation}\label{eq6_19_8}\begin{split}
E_{\text{Cas}}^{0,\text{XY}}
\sim &  -\frac{\hbar c R}{4\pi d^2}\sum_{s=0}^{\infty}\frac{(-1)^{(s+1)(\alpha_{\text{X}}+\alpha_{\text{Y}})}}{(s+1)^{2}}   \int_{\mu}^{\infty} dt \sqrt{t^2 -\mu^2}
   \exp\left( - 2 (s+1) t  \right)\\
  =& -\frac{\hbar c R}{8\pi d^2}\sum_{s=0}^{\infty}\frac{(-1)^{(s+1)(\alpha_{\text{X}}+\alpha_{\text{Y}})}}{(s+1)^{3}}\mu  K_{1}(2(s+1)\mu),
\end{split}\end{equation}
whereas the next-to-leading order term is
\begin{equation}\label{eq6_22_1}\begin{split}
E_{\text{Cas}}^{1,\text{XY}}
\sim &  -\frac{\hbar c R}{4\pi d^2} \sum_{s=0}^{\infty}\frac{(-1)^{(s+1)(\alpha_{\text{X}}+\alpha_{\text{Y}})}}{(s+1)^{2}}   \int_{\mu}^{\infty} dt \sqrt{t^2 -\mu^2}
  \mathcal{T}^{\text{X}}  \exp\left( - 2 (s+1) t  \right).
\end{split}\end{equation}
In the massless limit, $m=\mu=0$, and we find that
\begin{align*}
E_{\text{Cas}, m=0}^{0,\text{XY}}
\sim &  -\frac{\hbar c R}{4\pi d^2}\sum_{s=0}^{\infty}\frac{(-1)^{(s+1)(\alpha_{\text{X}}+\alpha_{\text{Y}})}}{(s+1)^{2}}   \int_{\mu}^{\infty} dt\, t
   \exp\left( - 2 (s+1) t  \right),
\end{align*}
\begin{align*}
E_{\text{Cas}, m=0}^{1,\text{XY}}
\sim &  -\frac{\hbar c R}{4\pi d^2} \sum_{s=0}^{\infty}\frac{(-1)^{(s+1)(\alpha_{\text{X}}+\alpha_{\text{Y}})}}{(s+1)^{2}}   \int_{0}^{\infty} dt\, t
  \mathcal{T}_0^{\text{X}}  \exp\left( - 2 (s+1) t  \right),
\end{align*}
where
\begin{align*}
\mathcal{T}_0^{\text{D}}=& \frac{\vep}{3},\\
\mathcal{T}_0^{\text{N}}=&\vep\left\{\frac{1}{3}- \frac{4}{3t}(s+1) \right\}.
\end{align*}
Direct integration and summation over $s$ using
\begin{equation}\label{eq6_19_7}
\begin{split}
\sum_{s=0}^{\infty}\frac{1}{(s+1)^k}=&\zeta(k),\\
\sum_{s=0}^{\infty}\frac{(-1)^{s+1}}{(s+1)^k}=&-\left(1-2^{1-k}\right)\zeta(k),
\end{split}
\end{equation} give
\begin{equation}\label{eq6_23_1}\begin{split}
E_{\text{Cas}, m=0}^{0, \text{DD}}=E_{\text{Cas}, m=0}^{0, \text{NN}}=&-\frac{\hbar c \pi^3 R}{1440 d^2},\\
E_{\text{Cas}, m=0}^{0, \text{DN}}=E_{\text{Cas}, m=0}^{0, \text{ND}}=&\frac{7\hbar c \pi^3 R}{11520 d^2},
\end{split}\end{equation}and
\begin{equation}\label{eq6_23_2}\begin{split}
E_{\text{Cas}, m=0}^{1, \text{DD}}=&E_{\text{Cas}, m=0}^{0, \text{DD}}\left(\frac{1}{3}\right)\frac{d}{R},\\
E_{\text{Cas}, m=0}^{1, \text{DN}}=&E_{\text{Cas}, m=0}^{0, \text{DN}}\left(\frac{1}{3}\right)\frac{d}{R},\\
E_{\text{Cas}, m=0}^{1, \text{ND}}=&E_{\text{Cas}, m=0}^{0, \text{ND}}\left(\frac{1}{3}-\frac{160}{7\pi^2}\right)\frac{d}{R},\\
E_{\text{Cas}, m=0}^{1, \text{NN}}=&E_{\text{Cas}, m=0}^{0, \text{NN}}\left(\frac{1}{3}-\frac{40}{\pi^2}\right)\frac{d}{R}.
\end{split}\end{equation}
which agree with the results \eqref{eq6_22_9} obtained in \cite{37,40}.

\begin{figure}[h]
\epsfxsize=0.4\linewidth \epsffile{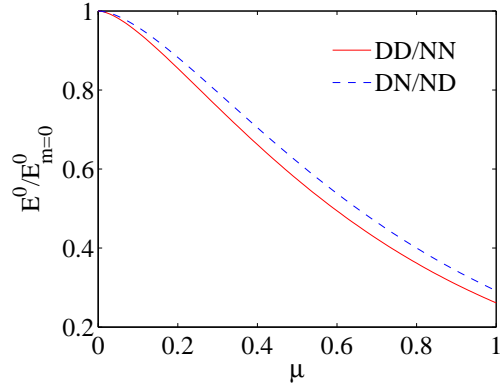}
 \caption{\label{f1} The dependence of $E_{\text{Cas}}^{0,\text{XY}}/E_{\text{Cas}, m=0}^{0,\text{XY}}$ on $\mu$. }\end{figure}
\begin{figure}[h]
\epsfxsize=0.4\linewidth \epsffile{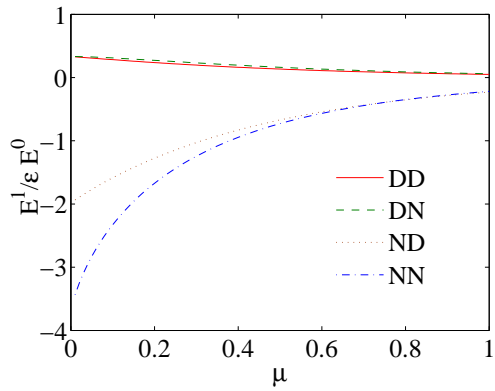}
 \caption{\label{f2} The dependence of $\vartheta^{\text{XY}}=E_{\text{Cas}}^{1,\text{XY}}/(\vep E_{\text{Cas}}^{0,\text{XY}})$ on $\mu$. }\end{figure}

 In Fig. \ref{f1}, we plot the leading term normalized by the massless leading term as a function of $\mu$. One observe that when the mass increases, the magnitude of the Casimir interaction energy decreases.

Define
 $$\vartheta^{\text{XY}}=\frac{E_{\text{Cas}}^{1,\text{XY}}}{\vep E_{\text{Cas}}^{0,\text{XY}}},$$  the ratio of the next-to-leading order term to the leading order term dividing by $d/R$, so that
 \begin{align*}
 E_{\text{Cas}}^{\text{XY}}\sim E_{\text{Cas}}^{0, \text{XY}}\left(1+\vartheta^{\text{XY}}\frac{d}{R}+\ldots\right).
 \end{align*}
 This is a good measure of the deviation from the proximity force approximation. In Fig. \ref{f2}, we plot the dependence of $\vartheta^{\text{XY}}$ on $\mu$. We observe that $\vartheta^{\text{DD}}$ and $\vartheta^{\text{DN}}$ are positive, while $\vartheta^{\text{ND}}$ and $\vartheta^{\text{NN}}$ are negative. Moreover, their magnitudes are decreasing functions of $\mu$. This implies that the correction to proximity force approximation becomes smaller when the mass is large.  

\section{Proximity force approximation}
The Casimir interaction energy density of massive scalar field between two parallel plates separated by a distance $d$ is given by
\begin{align*}
\mathcal{E}_{\text{Cas}}^{\parallel}(d) =&\frac{\hbar c}{4\pi^2}\int_0^{\infty} du \, u^2\ln\left\{1-(-1)^{\alpha_{\text{X}}+\alpha_{\text{Y}}}\exp\left(-2d\sqrt{u^2+\tilde{m}^2} \right)\right\},
\end{align*}where X, Y = D or N are the boundary conditions on the plates, $\alpha_{\text{D}}=0$, $\alpha_{\text{N}}=1$, and
$$\tilde{m}=\frac{mc}{\hbar}.$$
Using
the fact that
\begin{align*}
\int_0^R dr\, r f(R+d-\sqrt{R^2-r^2})=&\int_0^R dz z f(R+d-z)\\
\sim &d\int_1^{\infty} dv (R+d-dv)f(dv)\\
\sim & dR\int_0^{\infty} dv f(d(v+1)),
\end{align*}where $z=\sqrt{R^2-r^2}=R+d-dv$, we find that the
proximity force approximation to the Casimir interaction energy between a sphere and a plate is
\begin{align*}
E_{\text{Cas}}^{\text{PFA}}= &2\pi\int_0^{R}dr \,r \mathcal{E}_{\text{Cas}}^{\parallel}\left(R+d-\sqrt{R^2-r^2}\right)\\
=&2\pi d R\int_0^{\infty}dv   \mathcal{E}_{\text{Cas}}^{\parallel}\left(d(v+1)\right) \\
=& 2\pi d R\int_0^{\infty}dv \frac{\hbar c}{4\pi^2}\int_0^{\infty} du \, u^2\ln\left\{1-(-1)^{\alpha_{\text{X}}+\alpha_{\text{Y}}}\exp\left(-2d(v+1)\sqrt{u^2+\tilde{m}^2} \right)\right\}  \\
=&-\frac{\hbar cR}{4\pi}\sum_{s=0}^{\infty}\frac{(-1)^{(s+1)(\alpha_{\text{X}}+\alpha_{\text{Y}})}}{(s+1)^2}  \int_0^{\infty} du \, \frac{u^2}{\sqrt{u^2+\tilde{m}^2}} \exp\left(-2d(s+1) \sqrt{u^2+\tilde{m}^2} \right) \\
=&-\frac{\hbar cR}{4\pi d^2}\sum_{s=0}^{\infty}\frac{(-1)^{(s+1)(\alpha_{\text{X}}+\alpha_{\text{Y}})}}{(s+1)^2}  \int_{\mu}^{\infty} dt\,  \sqrt{t^2-\mu^2} \exp\left(-2 (s+1) t \right).
\end{align*}Here we have made the change of variables $t=d\sqrt{u^2+\tilde{m}^2}$. Compare to \eqref{eq6_19_8}, we find that the leading order term of the Casimir interaction energy indeed coincides with that derived using proximity force approximation.

\section{Small mass asymptotic expansion}
The formulas \eqref{eq6_19_8} and \eqref{eq6_22_1} give respectively the leading order term and the next-to-leading order terms of the Casimir interaction energy between a sphere and a plate. These formulas involve integration and summation and the dependence on mass is not obvious. To get a better picture on the dependence of the Casimir interaction energy on mass, we derive here the asymptotic expansions of these formulas  when the mass $m$ is small, or more precisely, when the dimensionless variable $\mu =mcd/\hbar$ is small.

Making the change of variables
$$u=\sqrt{t^2-\mu^2},$$ and using the formula
$$e^{-v}=\frac{1}{2\pi i}\int_{c-i\infty}^{c+i\infty} dz\,\Gamma(z) v^{-z},$$
we find that
\begin{align*}
&\sum_{s=0}^{\infty}\frac{1}{(s+1)^{2-\beta}}\int_{\mu}^{\infty} dt \frac{\sqrt{t^2 -\mu^2}}{t^{\alpha}}
    \exp\left(
 - 2 (s+1) t  \right)\\=&
  \sum_{s=0}^{\infty}\frac{1}{(s+1)^{2-\beta}}\int_{0}^{\infty}du \frac{u^2}{\left(u^2+\mu^2\right)^{\frac{\alpha+1}{2}}}
    \exp\left(
 - 2 (s+1) \sqrt{u^2+\mu^2} \right)\\=&\frac{1}{2\pi i}\int_{c-i\infty}^{c+i\infty} dz\Gamma(z)2^{-z}
 \sum_{s=0}^{\infty}\frac{1}{(s+1)^{z+2-\beta}}\int_{0}^{\infty}du \frac{u^2}{(u^2+\mu^2)^{\frac{z+\alpha+1}{2}}}\\
 =&\frac{1}{2}\frac{1}{2\pi i}\int_{c-i\infty}^{c+i\infty} dz\Gamma(z)2^{-z} \mu^{2-z-\alpha}\zeta_R(z+2-\beta)\frac{\Gamma\left(\frac{3}{2}\right)\Gamma\left(\frac{z+\alpha-2}{2}\right)}{\Gamma\left(\frac{z+\alpha+1}{2}\right)}.
      \end{align*}
Using the formula
$$\Gamma(z)=\frac{2^{z-1}}{\sqrt{\pi}}\Gamma\left(\frac{z}{2}\right)\Gamma\left(\frac{z+1}{2}\right),$$
we then find that
\begin{align*}
&\sum_{s=0}^{\infty}\frac{1}{(s+1)^{2-\beta}}\int_{\mu}^{\infty} dt \frac{\sqrt{t^2 -\mu^2}}{t^{\alpha}}
    \exp\left(
 - 2 (s+1) t  \right)\\=&\frac{1}{8}\frac{1}{2\pi i}\int_{c-i\infty}^{c+i\infty} dz\Gamma\left(\frac{z}{2}\right)\Gamma\left(\frac{z+1}{2}\right)\mu^{2-z-\alpha}\zeta_R(z+2-\beta)\frac{ \Gamma\left(\frac{z+\alpha-2}{2}\right)}{\Gamma\left(\frac{z+\alpha+1}{2}\right)}.
\end{align*}
Similarly, we have
\begin{align*}
&\sum_{s=0}^{\infty}\frac{(-1)^{s+1}}{(s+1)^{2-\beta}}\int_{\mu}^{\infty} dt \frac{\sqrt{t^2 -\mu^2}}{t^{\alpha}}
    \exp\left(
 - 2 (s+1) t  \right)\\=&-\frac{1}{8}\frac{1}{2\pi i}\int_{c-i\infty}^{c+i\infty} dz\Gamma\left(\frac{z}{2}\right)\Gamma\left(\frac{z+1}{2}\right)\mu^{2-z-\alpha}\left(1-2^{\beta-z-1}\right)\zeta_R(z+2-\beta)\frac{ \Gamma\left(\frac{z+\alpha-2}{2}\right)}{\Gamma\left(\frac{z+\alpha+1}{2}\right)}.
\end{align*}
From these, we can compute the small $\mu$ asymptotic expansions by taking residues at the poles of the integrands.
In particular, we find that
\begin{equation}\begin{split}
&\sum_{s=0}^{\infty}\frac{1}{(s+1)^2}\int_{\mu}^{\infty} dt \sqrt{t^2 -\mu^2}
    \exp\left(
 - 2 (s+1) t  \right)\\ \sim &\frac{1}{4}\zeta_R(4)-\frac{\mu^2}{4}\zeta_R(2)\left(1+2\psi(1)+2\frac{\zeta_R'(2)}{\zeta_R(2)}-2\log\mu\right)-\frac{\pi}{3}\mu^3\\&+\frac{\mu^4}{16}\left(
 \frac{5}{2}+2\psi(1)+2\log 2\pi-2\log\mu\right)+\frac{1}{4}\sum_{k=2}^{\infty}\frac{(-1)^{k}}{(k+1)!k}\pi^{\frac{3}{2}-2k}\Gamma\left(\frac{2k-1}{2}\right)\zeta_R(2k-1)\mu^{2k+2},
\\
&\sum_{s=0}^{\infty}\frac{(-1)^{s+1}}{(s+1)^2}\int_{\mu}^{\infty} dt \sqrt{t^2 -\mu^2}
    \exp\left(
 - 2 (s+1) t  \right)\\ \sim &- \frac{7}{32}\zeta_R(4)+\frac{\mu^2}{8}\zeta_R(2)\left(1+2\psi(1)+2\frac{\zeta_R'(2)}{\zeta_R(2)}+2\log 2-2\log\mu\right)\\&+\frac{\mu^4}{16}\left(
 \frac{5}{2}+2\psi(1)+2\log 2\pi-4\log 2-2\log\mu\right)\\&+\frac{1}{4}\sum_{k=2}^{\infty}\frac{(-1)^{k}}{(k+1)!}\pi^{\frac{3}{2}-2k}\left(2^{2k-1}-1\right)\Gamma\left(\frac{2k-1}{2}\right)\zeta_R(2k-1)\mu^{2k+2},
      \end{split}\end{equation}\begin{equation}\begin{split}
&\sum_{s=0}^{\infty} \int_{\mu}^{\infty} dt \frac{\sqrt{t^2 -\mu^2}}{t^2}e^{-2(s+1)t}\\\sim &\frac{\pi}{8}\mu^{-1} -\frac{1}{4}\left(-2+2\psi(1)+2\log 2\pi-2\log\mu\right)-\frac{\mu}{2\pi}\zeta_R(2)\\&-\frac{1}{2}\sum_{k=1}^{\infty}\frac{(-1)^k}{k!}\frac{1}{2k-1}\pi^{-2k-\frac{1}{2}}\Gamma\left(\frac{2k+1}{2}\right)\zeta_R(2k+1)\mu^{2k},
\\
&\sum_{s=0}^{\infty}(-1)^{s+1} \int_{\mu}^{\infty} dt \frac{\sqrt{t^2 -\mu^2}}{t^2}e^{-2(s+1)t}\\\sim & -\frac{1}{4}\left(-2+2\psi(1)+2\log 2\pi-4\log 2-2\log\mu\right)-\frac{3\mu}{2\pi}\zeta_R(2)\\
&-\frac{1}{2}\sum_{k=1}^{\infty}\frac{(-1)^k}{k!}\frac{2^{2k+1}-1}{2k-1}\pi^{-2k-\frac{1}{2}}\Gamma\left(\frac{2k+1}{2}\right)\zeta_R(2k+1)\mu^{2k},
\end{split}\end{equation}\begin{equation}\begin{split}
&\sum_{s=0}^{\infty}\frac{1}{s+1} \int_{\mu}^{\infty} dt \frac{\sqrt{t^2 -\mu^2}}{t}e^{-2(s+1)t}\\\sim &\frac{1}{2}\zeta_R(2)-\frac{\pi\mu}{4}\left(2-4\log 2-2\log\mu\right)-\frac{\mu^2}{4}\left(3+2\psi(1)+2\log 2\pi-2\log\mu\right)\\
&+\frac{1}{2}\sum_{k=1}^{\infty}\frac{(-1)^k}{k!}\frac{1}{(k+1)(2k+1)}\pi^{-2k-\frac{1}{2}}\Gamma\left(\frac{2k+1}{2}\right)\zeta_R(2k+1)\mu^{2k+2},
   \\
&\sum_{s=0}^{\infty}\frac{(-1)^{s+1}}{s+1} \int_{\mu}^{\infty} dt \frac{\sqrt{t^2 -\mu^2}}{t}e^{-2(s+1)t}\\ \sim &-\frac{1}{4}\zeta_R(2)+\frac{\pi\mu}{2}\log 2-\frac{\mu^2}{4}\left(3+2\psi(1)+2\log 2\pi-4\log 2-2\log\mu\right)\\
&+\frac{1}{2}\sum_{k=1}^{\infty}\frac{(-1)^k}{k!}\frac{2^{2k+1}-1}{(k+1)(2k+1)}\pi^{-2k-\frac{1}{2}}\Gamma\left(\frac{2k+1}{2}\right)\zeta_R(2k+1)\mu^{2k+2},
\end{split}\end{equation}\begin{equation}\begin{split}&
\sum_{s=0}^{\infty}\frac{1}{s+1} \int_{\mu}^{\infty} dt \frac{\sqrt{t^2 -\mu^2}}{t^3}e^{-2(s+1)t}\\\sim & -\frac{\pi}{8\mu}\left(1+4\log 2+2\log\mu\right)+\frac{1}{2}\left(2\psi(1)+2\log 2\pi-2\log\mu\right)+\frac{\mu}{2\pi}\zeta_R(2)\\
&+ \sum_{k=1}^{\infty}\frac{(-1)^k}{k!}\frac{1}{(2k-1)(2k+1)}\pi^{-2k-\frac{1}{2}}\Gamma\left(\frac{2k+1}{2}\right)\zeta_R(2k+1)\mu^{2k},\\
&\sum_{s=0}^{\infty}\frac{(-1)^{s+1}}{s+1} \int_{\mu}^{\infty} dt \frac{\sqrt{t^2 -\mu^2}}{t^3}e^{-2(s+1)t}\\\sim &-\frac{\pi}{4\mu} \log 2 +\frac{1}{2}\left(2\psi(1)+2\log 2\pi-4\log 2-2\log\mu\right)+\frac{3\mu}{2\pi}\zeta_R(2)\\
&+ \sum_{k=1}^{\infty}\frac{(-1)^k}{k!}\frac{2^{2k+1}-1}{(2k-1)(2k+1)}\pi^{-2k-\frac{1}{2}}\Gamma\left(\frac{2k+1}{2}\right)\zeta_R(2k+1)\mu^{2k}.
  \end{split}\end{equation}
Therefore, we find that up to terms of order lower than $\mu^4$, the small mass asymptotic expansion of the leading and next-to-leading order terms are given respectively by
\begin{equation}\label{eq6_24_3}
\begin{split}
E_{\text{Cas}}^{0, \text{DD/NN}}\sim &E_{\text{Cas}, m=0}^{0, \text{DD/NN}}\left\{1-\frac{15\mu^2}{\pi^2}\left(1+2\psi(1)+2\frac{\zeta_R'(2)}{\zeta_R(2)}-2\log\mu\right)-\frac{120}{\pi^3}\mu^3-\frac{45}{\pi^4}\mu^4\log\mu+O(\mu^4)\right\},\\
E_{\text{Cas}}^{0, \text{DN/ND}}\sim &E_{\text{Cas}, m=0}^{0, \text{DN/ND}}\left\{1-\frac{60\mu^2}{7\pi^2}\left(1+2\psi(1)+2\frac{\zeta_R'(2)}{\zeta_R(2)}+2\log 2-2\log\mu\right)+\frac{360}{7\pi^4}\mu^4\log\mu+O(\mu^4)\right\};
\end{split}
\end{equation}
\begin{equation}\label{eq6_24_4}
\begin{split}
E_{\text{Cas}}^{1,\text{DD}} \sim & E_{\text{Cas}, m=0}^{0, \text{DD}} \left\{\frac{1}{3}-\frac{15}{\pi^3}\mu+\mu^2\left(-\frac{5}{\pi^2}-\frac{60}{\pi^4}+\left[-\frac{10}{\pi^2}+\frac{60}{\pi^4}\right]\psi(1)-\frac{10}{\pi^2}\frac{\zeta_R'(2)}{\zeta_R(2)}
+\frac{60}{\pi^4}\log 2\pi+\left[\frac{10}{\pi^2}-\frac{60}{\pi^4}\right]\log\mu\right)\right.\\&\left.\hspace{2cm}-\frac{30}{\pi^3}\mu^3-\frac{15}{\pi^4}\mu^4\log\mu +O(\mu^4)\right\}\frac{d}{R},
\\
E^{1,\text{DN}}_{\text{Cas}}\sim &E_{\text{Cas}, m=0}^{0, \text{DN}} \left\{\frac{1}{3}+\mu^2\left(-\frac{20 }{7\pi^2}+\frac{480}{7\pi^4}+\left[-\frac{40 }{7\pi^2}-\frac{480}{7\pi^4}\right]\psi(1)-\frac{40}{7\pi^2}\frac{\zeta_R'(2)}{\zeta_R(2)}+
\left[-\frac{40 }{7\pi^2}+\frac{960}{7\pi^4}\right]\log 2\right.\right.\\&\left.\left.\hspace{2cm}-\frac{480}{7\pi^4}\log2\pi+\left[\frac{40 }{7\pi^2}+\frac{480}{7\pi^4}\right]\log\mu\right)-\frac{240}{7\pi^3}\mu^3+\frac{120}{7\pi^4}\mu^4\log\mu\right\}\frac{d}{R},
\\
E_{\text{Cas}}^{1,\text{ND}}\sim&E_{\text{Cas}, m=0}^{0, \text{ND}} \left\{\frac{1}{3}-\frac{160}{7\pi^2}+\frac{ 1440}{7\pi^3}\mu\log 2+\mu^2\left(-\frac{20 }{7\pi^2}-\frac{2400}{7\pi^4}+\left[-\frac{40 }{7\pi^2}-\frac{480}{7\pi^4}\right]\psi(1)-\frac{40}{7\pi^2}\frac{\zeta_R'(2)}{\zeta_R(2)}\right.\right.\\&\left.\left.\hspace{1cm}+
\left[-\frac{40 }{7\pi^2}+\frac{960}{7\pi^4}\right]\log 2-\frac{480}{7\pi^4}\log2\pi+\left[\frac{40 }{7\pi^2}+\frac{480}{7\pi^4}\right]\log\mu\right)+\frac{240}{7\pi^3}\mu^3+\frac{120}{7\pi^4}\mu^4\log\mu+O(\mu^4)\right\}\frac{d}{R},\\
E_{\text{Cas}}^{\text{NN}}\sim &E_{\text{Cas}, m=0}^{0, \text{NN}}\left\{\left(\frac{1}{3}-\frac{40}{\pi^2}\right)+\frac{15}{\pi^3}\mu\left(17-24\log 2-12\log\mu\right)+\mu^2\left(-\frac{5}{\pi^2}+\frac{300}{\pi^4}+\left[-\frac{10}{\pi^2}+\frac{60}{\pi^4}\right]\psi(1)\right.\right.\\&\left.\left.\hspace{2cm}-\frac{10}{\pi^2}\frac{\zeta_R'(2)}{\zeta_R(2)}
+\frac{60}{\pi^4}\log 2\pi+\left[\frac{10}{\pi^2}-\frac{60}{\pi^4}\right]\log\mu\right)-\frac{50}{\pi^3}\mu^3-\frac{15}{\pi^4}\mu^4\log\mu+O(\mu^4)\right\}\frac{d}{R}.
\end{split}
\end{equation}
From these expressions, it is obvious that when $\mu=0$, we recover the results \eqref{eq6_23_1} and \eqref{eq6_23_2} for the massless case. A somewhat surprising fact is that the dependence on $\mu$ is quite complicated. Naively, one would expect that the asymptotic expansions is in terms of even powers of $\mu$ only. However, as one can see from above,   we also have terms in odd powers of $\mu$ as well as  powers of $\mu$ times $\log\mu$ terms.

\begin{figure}[h]
\epsfxsize=0.4\linewidth \epsffile{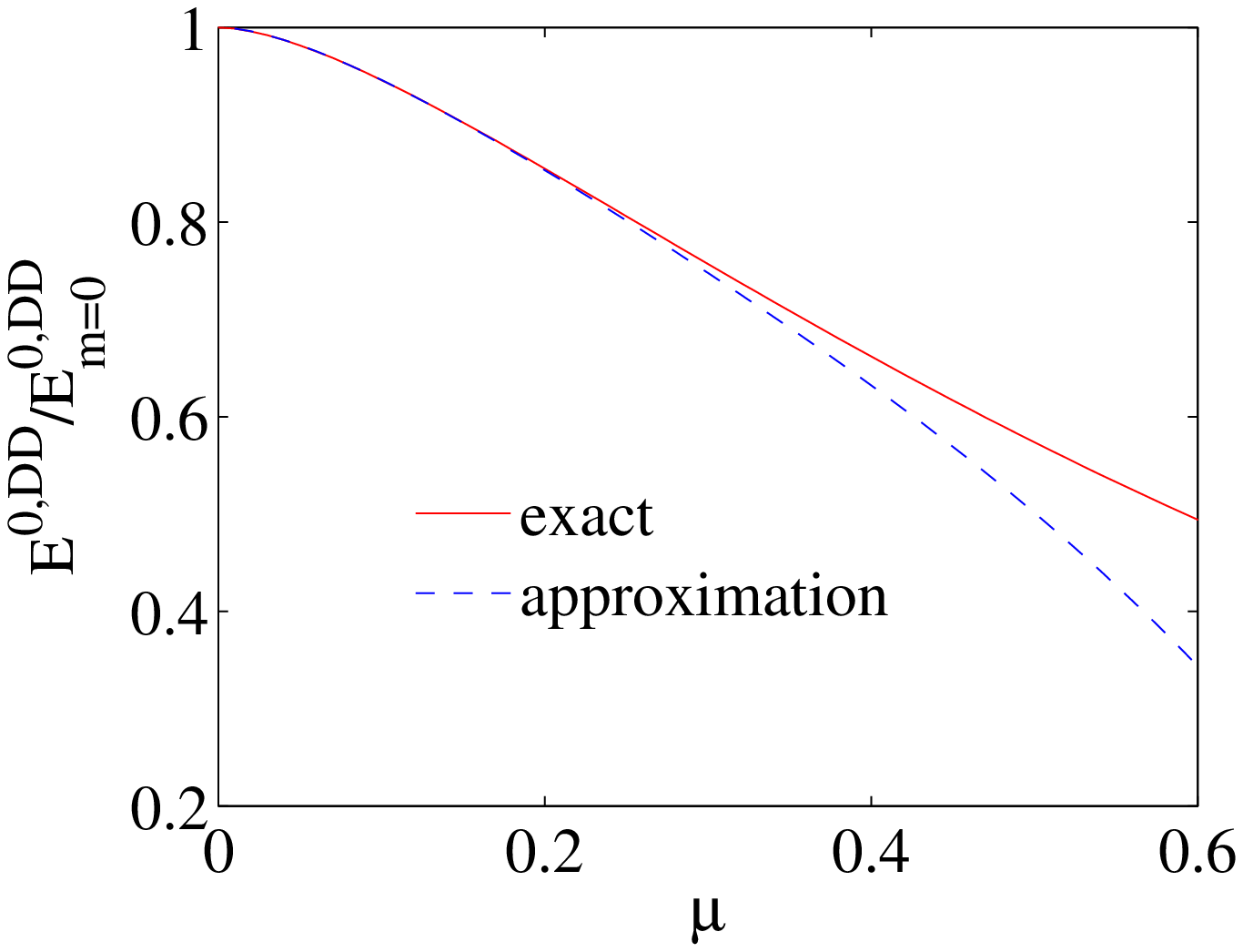}\epsfxsize=0.4\linewidth \epsffile{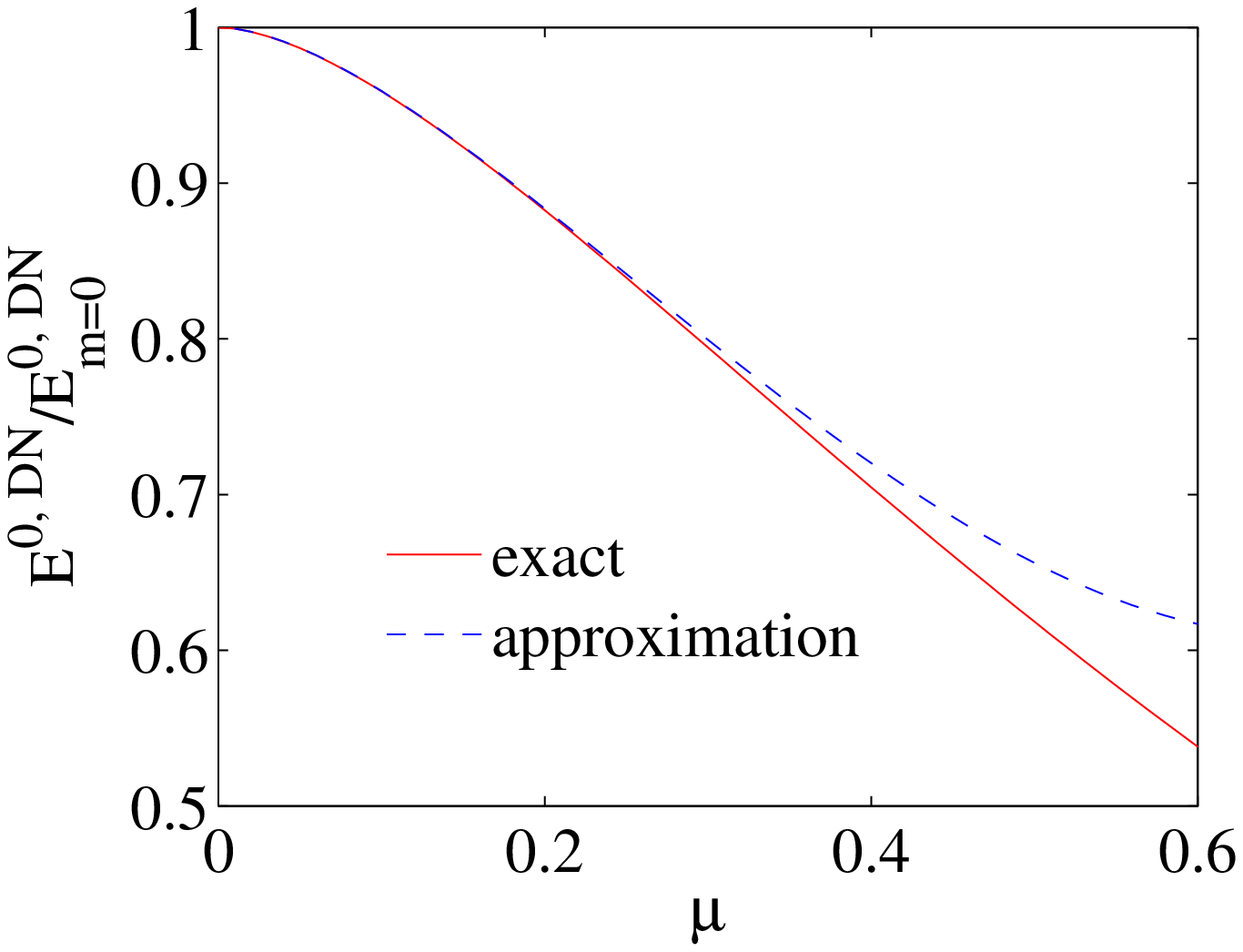}
 \caption{\label{f3} Comparison of  exact $E_{\text{Cas}}^{0,\text{XY}}/E_{\text{Cas}, m=0}^{0,\text{XY}}$ with the small mass approximation \eqref{eq6_24_3}. }\end{figure}
 \begin{figure}[h]
\epsfxsize=0.4\linewidth \epsffile{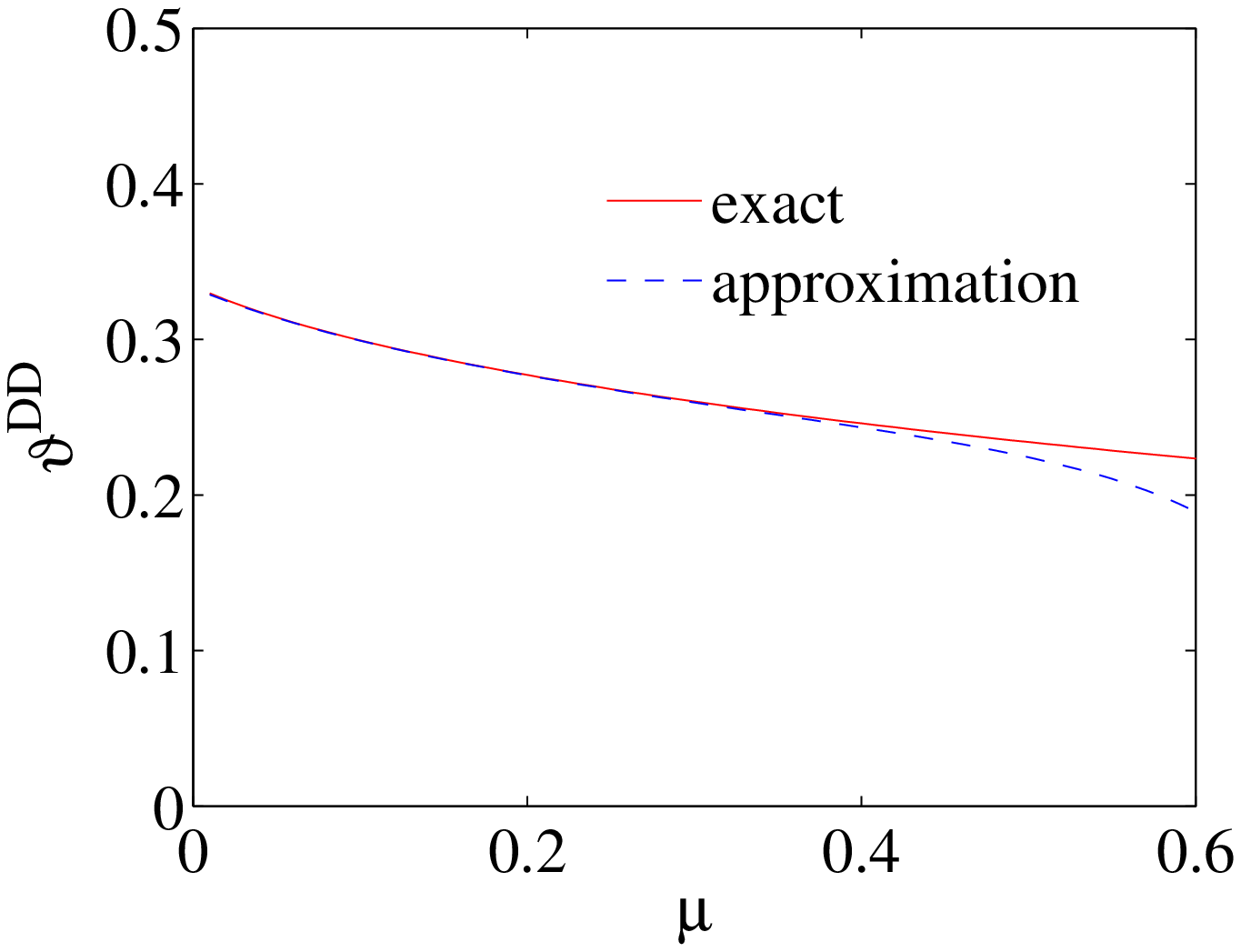}\epsfxsize=0.4\linewidth \epsffile{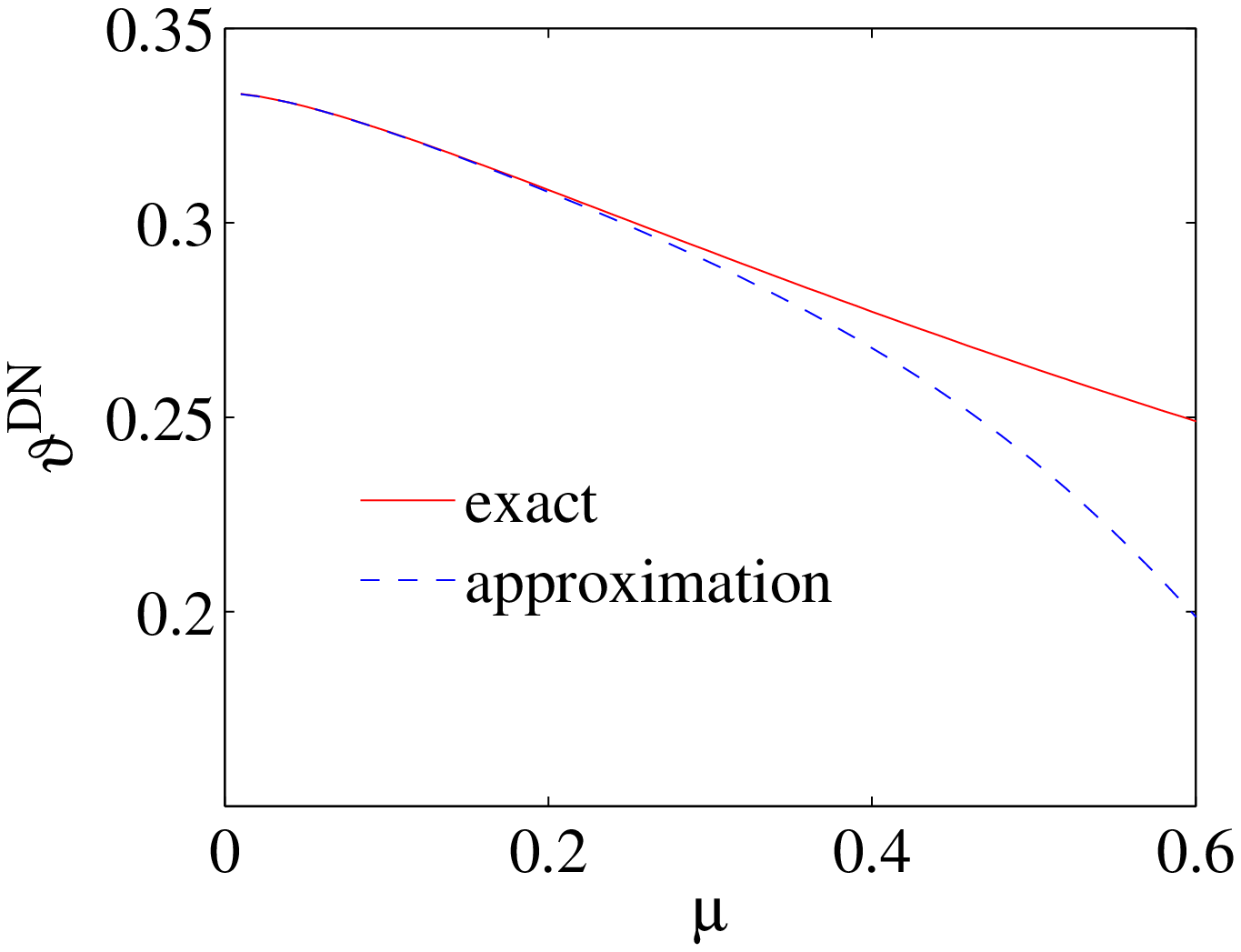}
 \caption{\label{f4} Comparison of  exact $\vartheta^{\text{XY}}$ with the small mass approximation \eqref{eq6_24_4}. }\end{figure}
 \begin{figure}[h]\epsfxsize=0.4\linewidth \epsffile{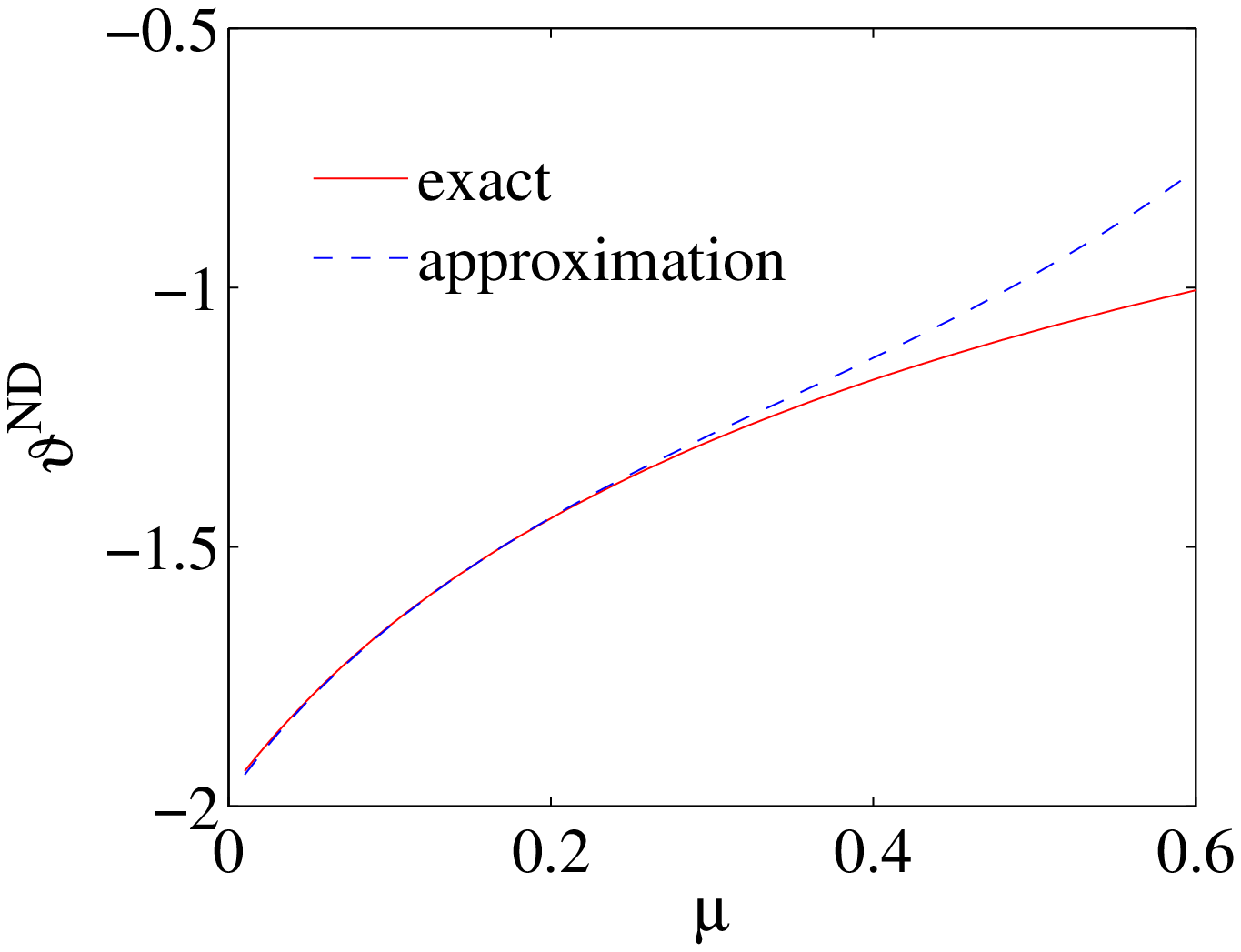}\epsfxsize=0.4\linewidth \epsffile{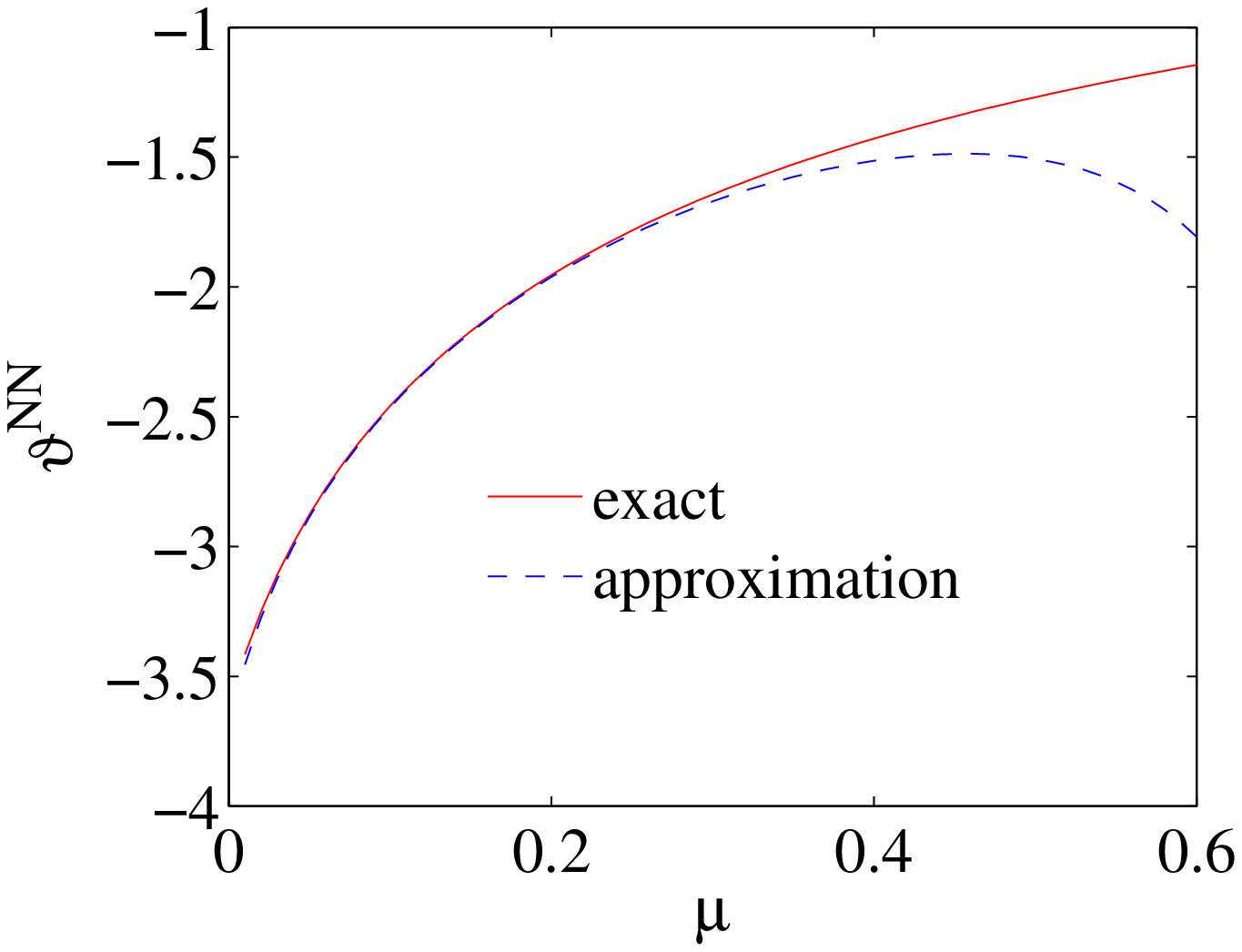}
 \caption{\label{f5} Comparison of  exact $\vartheta^{\text{XY}}$ with the small mass approximation \eqref{eq6_24_4}. }\end{figure}
 
 In Figs. \ref{f3}, \ref{f4} and \ref{f5}, we compare the exact values of $E_{\text{Cas}}^{0,\text{XY}}/E_{\text{Cas}, m=0}^{0,\text{XY}}$  and $\vartheta^{\text{XY}}$ to approximations given by the small mass expansions up to the terms of order $\mu^4\log\mu$  (\eqref{eq6_24_3} and \eqref{eq6_24_4}). We find that the small mass expansions \eqref{eq6_24_3} and \eqref{eq6_24_4} give  quite good approximations for $\mu\leq 0.3$. For better approximations, one should include higher order terms of $\mu$. 
 
\section{Conclusion}

We consider the Casimir interaction between a sphere and a plate due to vacuum fluctuations of massive scalar fields. Scrutinizing the derivation of the TGTG formula for the Casimir interaction energy in the massless case, it is straightforward to see that the Casimir interaction energy for the massive case can be directly obtained from the massless case by replacing $\omega/c$ by $\sqrt{(\omega/c)^2-(mc/\hbar)^2}$, where $\omega$ is the frequency, and $m$ is the mass. Although such simple relation exists for the TGTG formulas of the Casimir interaction energies, the dependence of the Casimir interaction energies on the mass of the scalar field is very complicated.

Since the Casimir interaction becomes significant when the separation between the sphere and the plate is very small, we derive the small separation asymptotic expansions of the Casimir interaction energies up to the next-to-leading order terms. The final results have nontrivial dependence on mass. Unlike the massless case, where the leading order terms and next-to-leading order terms can be expressed as simple rational functions of the distance, the leading order terms and next-to-leading order terms in the massive case can only be expressed as   infinite sums over some integrals. Thus, to understand the dependence of these terms on the mass of the field, we can perform some numerical computations or compute the asymptotic expansions of these terms when the mass is small. Naively, one would think that the small mass expansions only contain even powers of mass. However, after some tedious computations, we find that the small mass asymptotic expansions are actually quite complicated, as they contain odd powers of mass as well as powers of mass times logarithm of mass. These show that the dependence of the Casimir interaction on the mass of the scalar field is far more complicated and interesting.
\bigskip
\begin{acknowledgments}\noindent
 We would like to acknowledge the Ministry of   Education of Malaysia  for supporting this work under the  FRGS grant FRGS/1/2013/ST02/UNIM/02/2.
\end{acknowledgments}


\begin{thebibliography}{10}
\bibitem{15} H. B. G. Casimir, Proc. Kon. Nederland. Akad. Wetensch. B \textbf{51}, 793 (1948).
\bibitem{16} R.M. Cavalcanti, Phys. Rev. D \textbf{69}, 065015 (2004).
\bibitem{17} A. Lambrecht, P.A. Maia-Neto, and S. Reynaud, New Journal of Physics \textbf{8}, 243 (2006).
\bibitem{18} A. Bulgac, P. Magierski and A. Wirzba,   Phys. Rev. D \textbf{73}, 025007 (2006).





\bibitem{19} T. Emig, R. L. Jaffe, M. Kadar and A. Scardicchio,  Phys. Rev. Lett. \textbf{96}, 080403 (2006).


\bibitem{20} S. J. Rahi, T. Emig, R. L. Jaffe and M. Kardar, Phys. Rev. A \textbf{78}, 012104 (2008).

\bibitem{21} T. Emig,  N. Graham,  R. L. Jaffe   and M. Kardar, Phys. Rev. Lett. \textbf{99}, 170403 (2007).



\bibitem{22} T. Emig and R. L. Jaffe,  J. Phys. A: Math. Theor. \textbf{41}, 164001 (2008).

\bibitem{23} T. Emig, J. Stat. Mech. \textbf{0804}, P04007 (2008).




\bibitem{24} M. Bordag,   Phys. Rev. D \textbf{73}, 125018 (2006).

\bibitem{25} M. Bordag, Phys. Rev. D \textbf{75}, 065003 (2007).




\bibitem{26} O. Kenneth and I. Klich, Phys. Rev. Lett. \textbf{97}, 160401 (2006).


\bibitem{27} O. Kenneth and I. Klich,  Phys. Rev. B \textbf{78}, 014103 (2008).


\bibitem{28} K. A. Milton and J. Wagner,  J. Phys. A: Math. Theor. \textbf{41}, 155402 (2008).

\bibitem{29} K. Milton and J. Wagner, Phys. Rev. D \textbf{77}, 045005 (2008).

\bibitem{30} A. Wirzba,     J.  Phys. A  \textbf{41}, 164003 (2008).

\bibitem{31} T. Emig, N. Graham, R. L. Jaffe and M. Kadar, Phys. Rev. D \textbf{77}, 025005 (2008).

\bibitem{32} S. J. Rahi, T. Emig, N. Graham, R. L. Jaffe and M. Kadar, Phys, Rev. D \textbf{80}, 085021 (2009).
\bibitem{33} D. A. R. Dalvit, F. C. Lombardo, F. D. Mazzitelli and R. Onofrio, Phys. Rev. A \textbf{74}, 020101(R) (2006).



\bibitem{34} F. D. Mazzitelli, D. A. R. Dalvit and F. C. Lombardo, New. J. Phys. \textbf{8}, 240 (2006).

\bibitem{35} F. C. Lombardo, F. D. Mazzitelli, P. I. Villar and D. A. R. Dalvit, Phys. Rev. A \textbf{82}, 042509 (2010).
\bibitem{1}  L. P. Teo, Int. J. Mod. Phys. A \textbf{27}, 1230021 (2012).




\bibitem{37} M. Bordag and V. Nikolaev,  J. Phys. A: Math. Theor. \textbf{41}, 164002 (2008).
\bibitem{38} M. Bordag and V. Nikolaev,   Phys. Rev. D \textbf{81}, 065011 (2010).
\bibitem{39}  M. Bordag and I. Pirozhenko, Phys. Rev. D \textbf{81}, 085023 (2010).
\bibitem{40} L. P. Teo, M. Bordag and V. Nikolaev, Phys. Rev. D \textbf{84}, 125037 (2011).


\bibitem{41} L. P. Teo, Phys. Rev. D \textbf{84}, 025022 (2011).

\bibitem{42} L. P. Teo, Phys. Rev. D \textbf{84}, 065027 (2011).
\bibitem{43} L. P. Teo, Phys. Rev. D \textbf{85}, 045027 (2012).
\bibitem{44} L. P. Teo, Phys. Rev. D \textbf{88}, 045019 (2013).
\bibitem{45} L. P. Teo, Phys. Rev. A \textbf{89}, 052509 (2014).
\bibitem{46} L. P. Teo, J. Math. Phys. \textbf{55}, 043508 (2014).
\bibitem{47} L. P. Teo, Phys. Rev. D \textbf{89}, 105033 (2014).
\bibitem{48} L. P. Teo, Phys. Rev. D \textbf{90}, 045012 (2014).
\bibitem{49} L. P. Teo, Phys. Rev. D \textbf{91}, 085034 (2015).
\bibitem{50} L. P. Teo, Phys. Rev. D \textbf{91}, 125030 (2015).

\bibitem{3} M. Bordag, E. Elizalde, K. Kirsten, and S. Leseduarte, Phys. Rev. D \textbf{56}, 4896 (1997).
\bibitem{14} A. A. Saharian, Phys. Rev. D \textbf{63}, 125007 (2001).
\bibitem{13} A. A. Saharian, Phys. Rev. D \textbf{73}, 064019 (2006).
\bibitem{12} A. Chatrabhuti, P. Patcharamaneepakorn and P. Wongjun, J. High Energy Phys \textbf{08}, 019 (2009).
\bibitem{11} S. C. Lim and L. P. Teo, Ann. Phys. \textbf{324}, 1676 (2009).
\bibitem{10} G. Gazzola, M. C. Nemes, W. F. Wreszinski, Ann. Phys. \textbf{324}, 2095 (2009).
\bibitem{9} E. Elizalde, A. A. Saharian and T. A. Vardanyan, Phys. Rev. D \textbf{81}, 124003 (2010).
\bibitem{8} K. A. Milton and A. A. Saharian, Phys. Rev. D \textbf{85}, 064005 (2012).
\bibitem{7} J. M. Mu$\tilde{\text{n}}$oz Casta$\tilde{\text{n}}$eda, J. Mateos Guilarte and A. Moreno Mosquera, Phys. Rev. D \textbf{87}, 105020 (2013).
\bibitem{6} H. F. Mota and K. Bakke, Phys. Rev. D \textbf{89}, 027702 (2014).

\bibitem{5} S. Bellucci, A. A. Saharian, A. H. Yeranyan, Phys. Rev. D \textbf{89}, 105006 (2014).

\bibitem{4} A. A. Saharian and V. F. Manukyan, Classical Quantum Gravity \textbf{32}, 025009 (2015).
\bibitem{2} F. D. Mera and S. A. Fulling, J. Phys. A \textbf{48}, 45402 (2015).
\bibitem{51} I. S. Gradshteyn and I. M. Ryzhik, \emph{Table of integrals, series and products} (Academic Press, 2007).

\bibitem{52} M. Abramowitz and I. A. Stegun, \emph{Handbook of
Mathematical Functions} (Dover, New York, 1972).
\end{thebibliography}
\end{document}